\documentclass[pre,twocolumn,superscriptaddress,showpacs]{revtex4-1}

\usepackage{graphicx}
\usepackage{amsmath}
\usepackage{latexsym}
\usepackage{comment}
\usepackage{bm}
\usepackage{gensymb}
\usepackage{hyperref}
\hypersetup{colorlinks,urlcolor=blue}

\begin{document}
\title{Phyllotaxis, disk packing and Fibonacci numbers}
\date{\today}
\author{A. Mughal}
\affiliation{Institute of Mathematics, Physics and Computer Science, Aberystwyth University, Penglais, Aberystwyth, Ceredigion, Wales, SY23}
\author{D. Weaire} 
\affiliation{Foams and Complex Systems, School of Physics, Trinity College Dublin, Dublin 2, Ireland}

\begin{abstract}
We consider the evolution of the packing of disks (representing the position of buds) that are introduced at the top of a surface which has the form of a growing stem. They migrate downwards, while conforming to three principles, applied locally: dense packing, homogeneity and continuity. We show that spiral structures characterised by the widely observed Fibonacci sequence (1,1,2,3,5,8,13...), as well as related structures, occur naturally under such rules. Typical results are presented in a animation. 
\end{abstract}

\maketitle


\section{Introduction}

\begin{figure} 
\begin{center}
\centering
\includegraphics[width=0.325\columnwidth ]{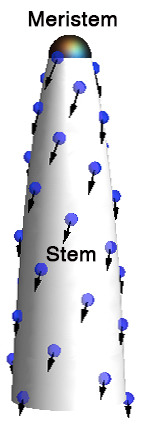}
\caption{In a growing plant shoot, which has a constant profile, new buds (blue) emerge at the tip of the shoot (the meristem). The buds move down the stem - in the frame of reference in which the profile is fixed - as indicated by the arrows. The position of the buds may rearrange before settling down to their eventual relative location, as the diameter tends to a constant.}
\label{fig1}
\end{center}
\end{figure}

Phyllotaxis (the arrangement of buds or branches on a stem, or petals on a flower) has long been debated \cite{airy1872leaf, hofstadter2013alan}, particularly with regard to the widespread occurrence of spiral structures that are related to the Fibonacci sequence \cite{levitov1991fibonacci, levitov1991energetic, douady1992phyllotaxis, atela2002dynamical, Pennybacker:2013kp}.  Here we offer a theoretical model which relates the problem to disk packings, extending previous work \cite{Mitchison:1977uy, van1907mathematische} that seeks explanations in that way.

We consider a tapered shape roughly representative of a plant stem, see Fig (\ref{fig1}). Buds, steadily introduced at top (the meristem), travel downwards with respect to the growing stem, continuously adapting their arrangement to the local diameter. In what follows, the buds (blue dots) are represented by the centres of hard disks. Three simple principles are adduced to dictate the evolving packing of these disks, as they migrate downwards, taking account of the changing ratio of their fixed diameter to that of the stem:
 
\vspace{4mm}
\fbox{\parbox{7.6cm}{
\begin{enumerate}
\item {\it Maximal packing}. Locally, disks are to remain as densely packed as possible.
\item {\it Homogeneity}. The local arrangement of disks is homogeneous - that is, without the defects that are found in the maximal packings for cylinders
\item {\it Continuity}. No abrupt finite change of structure is allowed.
\end{enumerate}
	}
}
\vspace{5mm}

We here offer no particular rationalisation of these elementary principles, in biological terms. In as much as they prove to be successful, this is an interesting challenge that deserves further experiment and observation.    

Our method is to adapt results for the dense packing of hard disks on a cylinder \cite{Mughal:2011tg, Mughal:2012jd,  Mughal:2014bk}, where helical symmetry arises naturally. The procedure is related to previous ideas in which phyllotaxis is described in terms of the close packing of disks \cite{Mitchison:1977uy, van1907mathematische}; this provides a conceptually simple and explicit model, capable of wide variation. The most striking feature of the results is the occurrence of structures related to the Fibonacci series (1,1,2,3,5,8,13...).

At this stage we maintain the model in its simplest form, without any variation of disk size or stem profile with time. The transformation that we will employ in order to realise results of the model is approximate - but any simple model based on the packing of circular disks on a surface of varying Gaussian curvature is necessarily approximate.

\section{Phyllotactic Notation}
The optimal packing of disks of diameter $d$ on a right circular cylinder of diameter $D$ can be achieved by ``rolling up'' the 2D hexagonal close-packing structure only for special values of $D/d$, these we call \emph{symmetric structures} \cite{Mughal:2011tg, Mughal:2012jd,  Mughal:2014bk}. For intermediate values the dense packing of disks on cylinders can be achieved by a affine shear of the symmetric structures, which we call \emph{rhombic disk packings}. These arrangements are described below.

\begin{figure*}
\begin{center}
\centering
\includegraphics[width=2.0\columnwidth ]{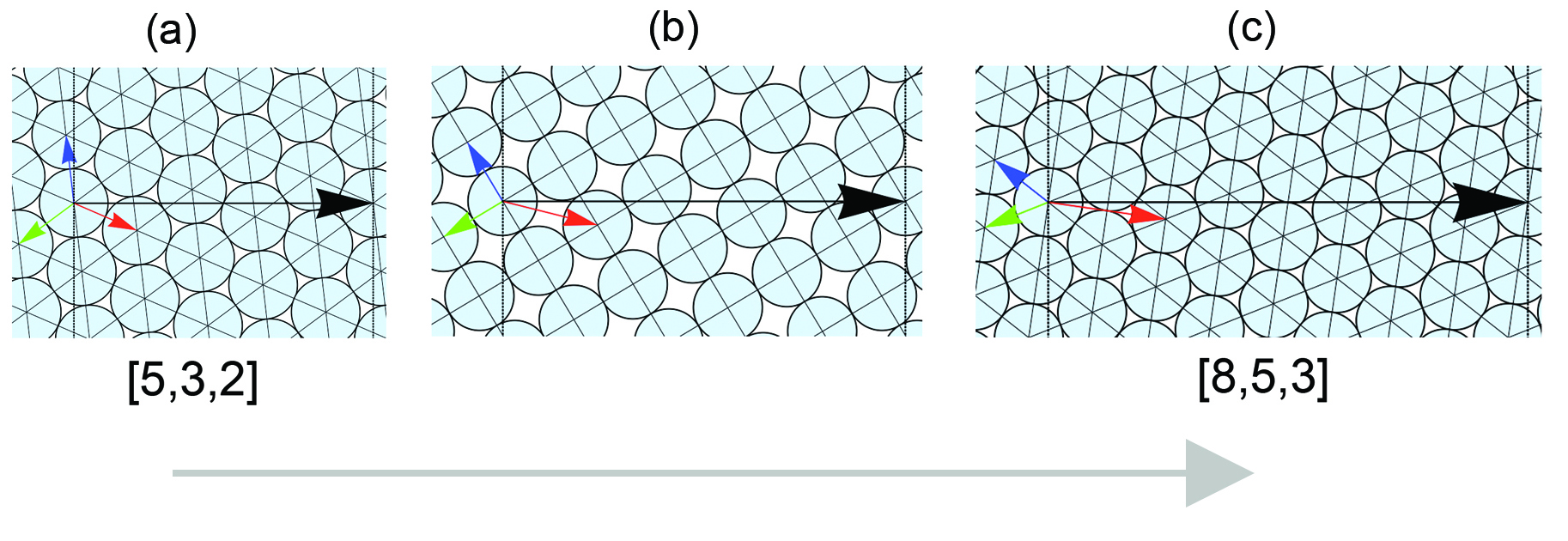}
\caption{
Arrangements of disks on a plane that can be excised and wrapped onto a cylinder of an appropriate diameter. The red, blue and green arrows indicate the directions of the  $\widehat{\bf a_1}$,  $\widehat{\bf a_2}$, and  $\widehat{\bf a_3}$ primitive lattice vectors, respectively. The black arrow indicates the periodicity vector. As shown, disks are arranged into a triangular lattice. The number of rows crossing the periodicity  vector in each of the three directions are given by the indices $l, m$ and $n$. The  first structure (a), labelled $[l=5, m=3, n=2]$ is a ``symmetric'' structure whereby the disks are arranged into an equilateral triangular lattice. Maximal density (without defects) is maintained, as the cylinder diameter is varied, by losing the contacts in the $\widehat{\bf a_1}$ direction - as shown in (b). The progression from the symmetric $[5,3,2]$ structure to the new symmetric arrangement $[8,5,3]$, as shown in (c), is through a series of such intermediate rhombic structures (grey arrow). 
}
\label{fig2}
\end{center}
\end{figure*}

\subsection{Symmetric structures}

A symmetric triangular lattice is one in which the disk centres form an equilateral triangular lattice. It can be wrapped onto a cylinder of the appropriate diameter if we can identify a periodicity vector ${\bf V}$ between a pair of lattice points, shown by the black arrow in Fig (\ref{fig2}a). The dotted lines, perpendicular to the periodicity vector, at the base and the head of ${\bf V}$ define a section that can be excised and wrapped around a cylinder of diameter $|{\bf V}|/\pi$ \cite{Mughal:2011tg, Mughal:2012jd,  Mughal:2014bk}.

A cylindrical pattern created in this way consists, in general, of spiral lines in three directions (corresponding to the primitive lattice vectors); exceptional cases include the limiting case of lines that go around the circumference or are parallel to the cylinder axis. Each of these three directions may be associated with a phyllotactic index (for which the traditional term in biology is \emph{parastichy}) - see Fig (\ref{fig2}) and \cite{Mughal:2014bk} for full details.

\subsection{Rhombic disk packings}

For intermediate values of the cylinder diameter the required seamless mapping can be accommodated by an affine transformation (or strain) of the lattice. Such a structure has maximal density, if defective structures are disallowed, as here \cite{Mughal:2011tg, Mughal:2012jd,  Mughal:2014bk}.

Fig (\ref{fig2}) illustrates one such transformation, beginning with the symmetric triangular disk packing $[5,3,2]$, where as indicated by the heavy black triangle all the angles of the lattice are the same (see Fig (\ref{fig2}a)). The lattice can be strained by an affine deformation that results in a homogenous \emph{rhombic} structure in which contacts in one direction become separated, while the others are maintained. An example is shown in Fig (\ref{fig2}b) whereby contacts are lost in the $\widehat{\bf a_1}$ direction. The final symmetric triangular disk packing $[8,5,3]$, shown in  Fig (\ref{fig2}c), is arrived at by a continuous series of intermediate rhombic structures - each of which can be wrapped onto the surface of a cylinder of the appropriate diameter.

In this manner the strained structure may proceed from one symmetric triangular disk packing to another as $\bf{V}$ is varied. Whenever a symmetric structure is encountered there is more than one possibility for such continuation. The choice which maintains maximum density is described in the next section. As reported previously \cite{Mughal:2011tg, Mughal:2012jd,  Mughal:2014bk}  - the three choices are the following:
\begin{eqnarray}
\left[m+n, m, n \right]
&\rightarrow&
\left[m, m-n, n \right]
\nonumber
\\
\left[m+n, m, n \right]
&\rightarrow&
\left[2m+n, m+n, n \right]
\nonumber
\\
\left[m+n, m, n \right]
&\rightarrow&
\left[m+2n, m+n, n \right],
\nonumber
\end{eqnarray}
where the new phyllotactic indices may be rearranged into descending order, according to convention.

\begin{figure*} 
\begin{center}
\centering
\includegraphics[width=1.65\columnwidth ]{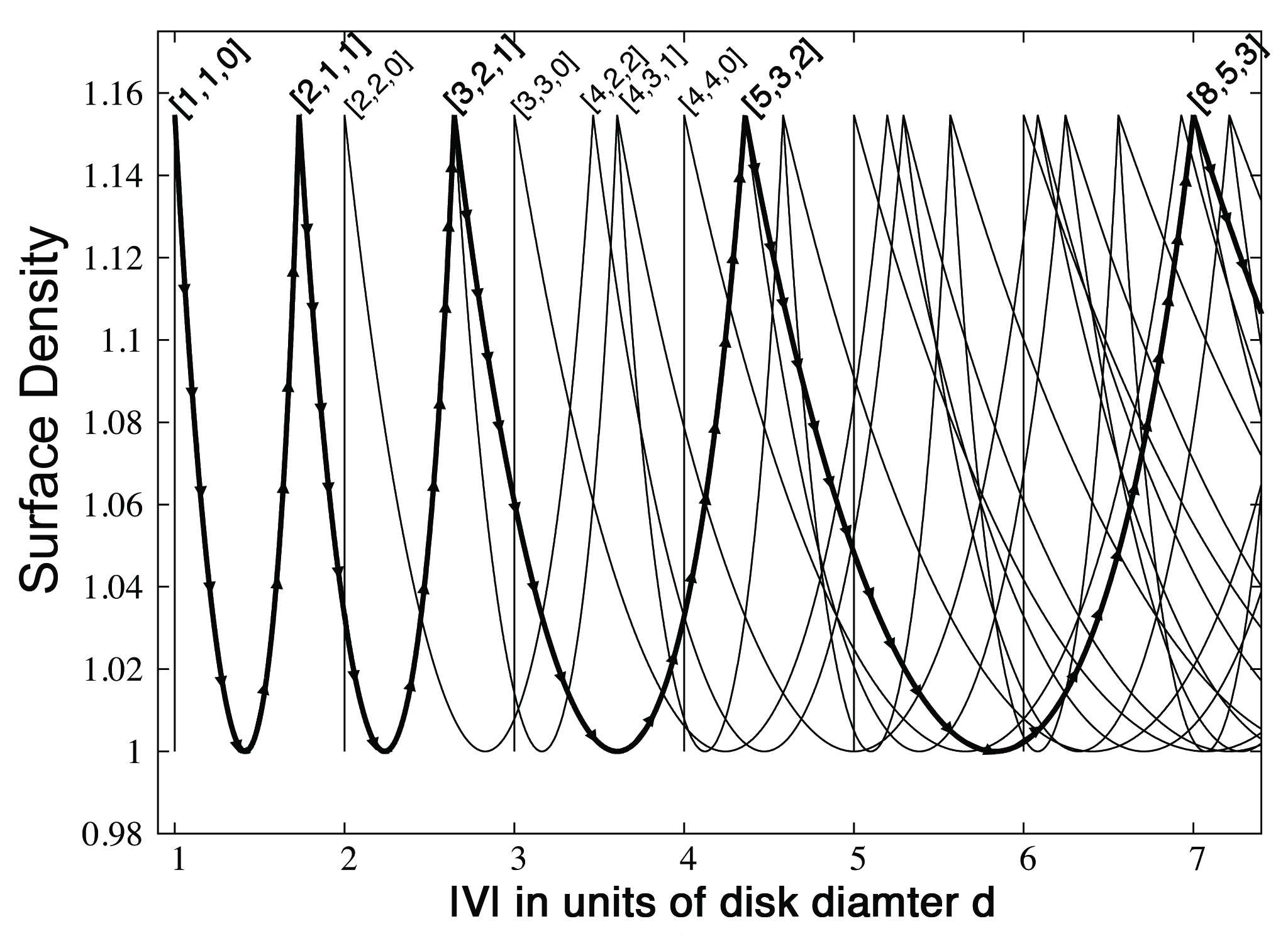}
\caption{Surface density of disk packings on the plane which are consistent with wrapping disks of unit diameter onto a cylinder of diameter $D=|{\bf V}|/\pi$. The length of the periodicity vector $|{\bf V}|$ is in units of the disk diameter $d$. Symmetric structures labeled $[l,m,n]$ correspond to points of maximum density. The bold curve is a sequence which accords with the stated principles starting from the symmetric structure $[1,1,0]$. The sequence of symmetric structures corresponding to the Fibonacci sequence are highlighted in bold.}
\label{fig3}
\end{center}
\end{figure*}

\section{Fibonacci Sequence}

Previous analytical results for the density of these disk packings are shown in Fig (\ref{fig3}). Note that the heavy line indicates the continuous sequence of structures that have maximal density, beginning with symmetric $[1,1,0]$ structure (as discussed below). 

Whenever we reach a cylinder diameter that is consistent with symmetric close packing (six contacts), of type $[l,m,n]$, there is more than one choice for further evolution - see Fig (\ref{fig3}): we may lose contacts in either the $\widehat{\bf a_1}$ or $\widehat{\bf a_3}$ directions. Maximal density (the first principle) is maintained by losing the contact in the $\widehat{\bf a_1}$ direction, this leads to the following progression, 

\begin{equation}
\left[ l,m,n \right]
\rightarrow
\left[ l',m',n' \right]
\end{equation}
where $l'=l+m$, $m'=l$ and $n'=m$. This is precisely the rule of progression in the Fibonacci sequence. 

\section{Projection onto a stem}

The above results can be used to create a continuously evolving pattern on a model stem, of the kind that has been stipulated. An appropriate mathematical projection (described below) of the cylindrical packings on to the stem slightly distorts the disks. 

\subsection{Stem profile}

We here consider an arbitrary but reasonable choice for the stem profile: a surface of revolution whose diameter is, given by, 
\begin{equation}
\mathcal{D}(\mathcal{Z})=\mathcal{D}_0 \tanh \left( \frac{\mathcal{Z}}{\mathcal{Z}_0} \right),
\label{eq:profile}
\end{equation}
where, at any height $\mathcal{Z}$, the ratio of stem and disk diameters is given by the fraction $\mathcal{D}(\mathcal{Z})/d$. Here $\mathcal{D}_0$ is the limiting diameter of the stem when $\mathcal{Z}\rightarrow \infty$, while $\mathcal{Z}_0$ is an arbitrary constant that can be tuned to give either a rapidly ($\mathcal{Z}_0 \rightarrow 0$) or slowly ($\mathcal{Z}_0 \rightarrow \infty$) changing diameter at the stem tip. If $\mathcal{Z}_0$ is too small it can lead to a very rapidly changing diameter at the stem tip making the projection (described below) unstable. 

The shape of the stem profile determines the precise details of the projection. In the simplest case, which we consider here, we choose the originating structure at the tip of the stem to be given by the $[1,1,0]$ symmetric arrangement. This is the simplest disk packing that can be wrapped on to a cylindrical surface and has a diameter given by $d/\pi$. Thus
at the tip (i.e. at $\mathcal{Z}=0$) we fix this to be the diameter of the stem. Above this point (i.e., at a height greater than $h$; see Fig (\ref{fig4})) we cut-off the surface since it is not possible to define a disk packing with a smaller diameter - the continuation of the surface, representing the meristem, is shown in grey. 

\subsection{The projection}

\begin{figure}
\begin{center}
\includegraphics[width=1.0\columnwidth ]{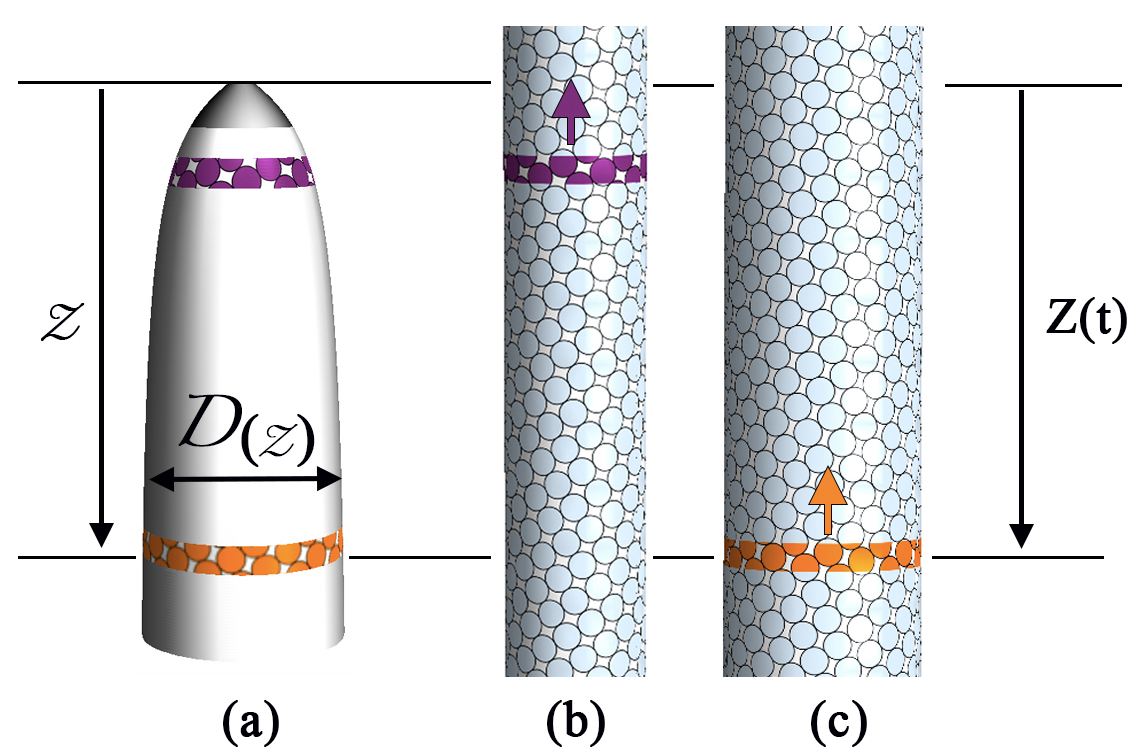}
\caption{(a) A stem of varying diameter $\mathcal{D}(\mathcal{Z})$. We cut off the stem at the tip where the diameter is $d/\pi$, since it is not possible to define a disk packing with a smaller diameter. The continuation of the surface (i.e. the meristem) is shown in dark grey for any height greater than $h$. At any point we can consider an infinitesimal slice of the stem: here two such slices are shown in orange and purple. The required section is taken from the corresponding cylindrical patterns of the appropriate diameter, as shown in (b) and (c). The excised cylindrical section moves upwards at a constant velocity (as indicated by the orange and purple arrows) so that at $t>0$ the transposed section is located higher up along the cylinder.}
\label{fig4}
\end{center}
\end{figure}

Associated with each point on the bold curve in Fig (\ref{fig3}) is a particular pattern on a cylinder of diameter $D$, described above and in Refs. \cite{Mughal:2011tg, Mughal:2012jd,  Mughal:2014bk}. 

Any one of these patterns, can be rotated around the axis of the cylinder (i.e., in the direction of the periodicity vector). In following this sequence we keep a particular point (disk centre) on the cylinder fixed at $z=0$ so that the variation of the structure with $D$ is continuous. This is equivalent to stipulating that the originating structure at the stem tip - in this case the $[1,1,0]$ packing - cannot rotate. 

From this condition the structure may then be projected on to a stem of varying diameter $\mathcal{D}(\mathcal{Z})$, as follows, in order to create an evolving pattern in accordance with the stated principles. Consider an \emph{infinitesimal} slice of the stem as shown in Fig (\ref{fig4}a), where the diameter $\mathcal{D}(\mathcal{Z})$ of the slice depends on the height $\mathcal{Z}$. We can excise a slice from a cylindrical pattern of the same diameter as the slice (bold curve in Fig (\ref{fig3})) and transpose it to the corresponding section of the stem. Note, (as mentioned above) since the azimuthal angle of the originating structure (i.e., the first slice) is fixed this implies that angle of each subsequent slice is also fixed (i.e., there is no $\mathcal{Z}$ dependent shift of the pattern when projected on to the stem). 

The process is illustrated by the two examples shown in Fig (\ref{fig4}b) and Fig (\ref{fig4}c), where the slice towards the tip with the smaller diameter (purple) is taken from the first cylinder and the the slice with the larger diameter (orange) is taken from the second cylinder. We are free to choose from where on the cylinder the slice is taken, since for any height $z(t)$ the cylinder has a constant diameter. 

To create a projection that can evolve over time the required region is mapped by a Galilean transformation, from the cylinder (moving frame) and on to the stem (fixed frame). 

The position of a given slice on the stem is fixed at a height $\mathcal{Z}$, as determined by Eq. ($\!\!$~\ref{eq:profile}). However, the position at which the slice is excised from the cylinder varies with time: it moves upwards along the length of the cylinder at a constant velocity. The required transformation for time $t\geq 0$ is,
\begin{equation}
\mathcal{Z}=z(t)+v_0\frac{\mathcal{D}_0}{\mathcal{D}(\mathcal{Z})}t,
\end{equation}
where $v_0$ is the limiting velocity as $\mathcal{Z}\rightarrow \infty$. At $t=0$ the two coordinate systems are coincident so that as shown in Fig (\ref{fig4}) the slice from the cylindrical pattern is level with the corresponding slice on the stem.

\subsection{Simulations}

\begin{figure*}
\begin{center}
\includegraphics[width=1.25\columnwidth ]{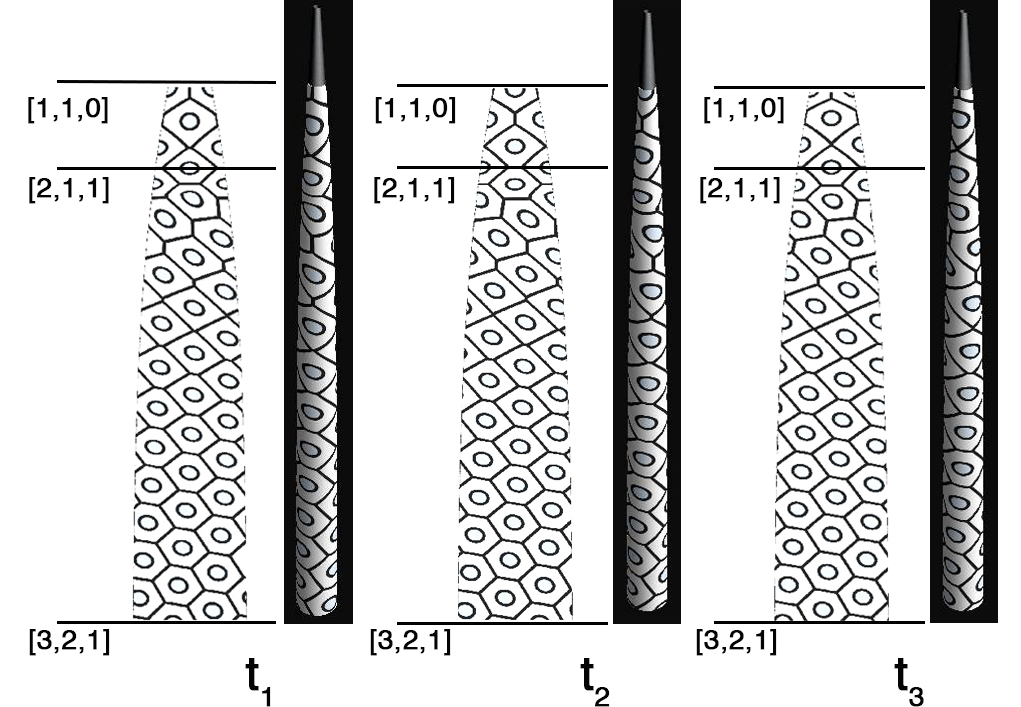}
\caption{Each pair of images is for a certain time ($t_1<t_2<t_3$). The full sequence of images can be viewed in the accompanying animation (see supplementary material). For each pair, the image on the right is the computationally generated projection on the stem. For graphical purposes we do not show the full disks. Instead we show the Voronoi tessellation (black lines) about the disk centres, highlighting nearest neighbour relationships. We also show the continuation of the surface (in dark grey) which is representative of the meristem. The accompanying images to the left show the projection unrolled on to a plane (as described in the text). The horizontal lines indicate the heights at which the pattern assume the symmetric structures  $[1,1,0]$ $[2,1,1]$ and $[3,2,1]$. To view the animation see \cite{movie}.}
\label{fig5}
\end{center}
\end{figure*}

In practice, the computational realisation of the projection described above uses a series of thin slices of \emph{finite} thickness $d{\mathcal{Z}}$. By stacking a series of cylindrical shells, of smoothly varying diameter, we can assemble a surface of revolution that approximates the desired projection at time $t$. 

To define the projection we need to specify the diameter at the tip of the stem ($\mathcal{Z}=0$) and at the bottom of the stem ($\mathcal{Z}\rightarrow \infty$). Here we set the diameter at the top to be given by $\pi \mathcal{D}(0)=d$, which corresponds to the length of the periodicity vector of the symmetric $[1,1,0]$ structure.  At the bottom the diameter is set to  $\pi \mathcal{D}(\infty)=\sqrt{7}d$, which corresponds to the length of the periodicity vector of the symmetric $[3,2,1]$ structure.

The computationally generated projection is shown in Fig (\ref{fig5}), where we show the results at three successive time steps ($t_1<t_2<t_3$). The full sequence of images is shown in the accompanying animation (described below) - to view the animation see \cite{movie}. 

For graphical purposes we do not show the full disks; instead we include the Voronoi tessellation about the disk centres (where the tessellation is generated about the disk centres on a given cylinder, this along with the disks is then projected on to the stem using the method described above). This is particularly helpful in visualising the rearrangement of disks as they move down the stem. Also shown (in dark grey) at the tip of the stem is the meristem represented by the continuation of the surface given by Eq. ($\!\!$~\ref{eq:profile}).

To allow the entire projection to be observed on the page we may ``unroll'' it onto the plane. That is: if a point on the stem is located in cylindrical polar coordinates at $(\mathcal{D}(\mathcal{Z})/2,\theta, \mathcal{Z})$ - where $\theta$ is the azimuthal angle - then it is mapped to the plane with Cartesian coordinates $u=(\mathcal{D}/2)\theta$ and $v=\mathcal{Z}$. These unrolled representations are shown to the left of each projection at any given time in Fig (\ref{fig5}) and in the accompanying animation. Note, that since the stem does not have vanishing Gaussian curvature the mapping on to a plane generates distorts the pattern; these effects are particularly pronounced towards the top. 

The advantage of unrolling the projection is that all key features can be easily identified. Of particular importance are the heights labelled $[1,1,0]$, $[2,1,1]$ and $[3,2,1]$ in Fig (\ref{fig5}), where the circumference of the stem is equal to the length of the periodicity vector of the specified symmetric structure. Between these heights the projection evolves from one symmetric arrangement to another by an affine shear of the type describe above. In terms of the Voronoi representation these transitions between symmetric structures involve  a topological rearrangement (T1) in which cells switch neighbours by the shrinkage of a Voronoi boundary. 

\subsection{Animation}

The accompanying animation is comprised of a sequence of stills at successive times  concatenated together. Each frame is separated from the next by a short time. In understanding the sequence, it is helpful to observe the progress of a disk as it emerges from the tip (meristem) and travels down the stem. 

At the tip, where the stem is narrowest, disks emerge individually and from alternating sides. In the system of classification presented here this arrangement corresponds to the $[1,1,0]$ structure - which is the simplest possible periodic arrangement of disks that can be wrapped onto a cylinder. The structure is commonly observed in plants whereby adjacent buds are offset by 180\degree along the stem; in the language of botany such patterns are known as \emph{distichous} (alternate) phyllotaxis. 

While the originating structure at the tip is the $[1,1,0]$ arrangement, moving down from the tip the diameter of the stem increases rapidly which forces the arrangement of disks to evolve. In these subsequent - more complex structures - the buds form helical arrangements (i.e., the angle between successive buds is less than 180\degree), which in botany are termed spiral phyllotaxis. Below the tip, the transition towards the symmetric $[2,1,1]$ structure is accomplished by a T1 rearrangement of the disk centres. In terms of the Voronoi diagram, edges in the direction of the stem gradually disappear and are replaced by edges perpendicular to the direction of the stem, which then gradually grow in length. 

Over the final (and longest) section of the stem the diameter increases slowly until towards the bottom it is almost constant. In this region the arrangement of disks gradually evolves from a $[2,1,1]$ to a $[3,2,1]$ arrangement. Again this transition, from a simpler to more complex structure, is accompanied by T1 transitions, whereby the edges that are parallel to the stem shrink to be replaced by edges that are perpendicular to it.

\section{Conclusions}

It may seem redundant to offer yet another elucidation of phyllotaxis when so much has gone before, including connections with disk packings, but the transparency and the automatic nature of the present model should commend it. Its precise relation to biological reality remains to be explored. Detailed observation of phyllotactic development are surprisingly sparse and more would be welcome for comparison with models such as this.  

This is perhaps all the more pressing since it is only relatively recently (considering the age of the subject) that the hormone auxin has been identified as the driver of phyllotaxis \cite{Reinhardt:2003}. As such there is now a well-established programme to model in detail the mechanical-biochemical process involved and its relation to optimal packing \cite{Pennybacker:2013kp, Newell:2008} (also see \cite{Pennybacker:2015} for a recent review article).

The present model may prove useful in this debate. An important test case may be  \emph{Agave parryi}, which has only lately been brought to the attention of the mathematical community \cite{Rivier:2006}: normally a spherical cactus with a $[13, 8, 5]$ structure but which in its final moments sprouts a huge seed-bearing mast of diminishing diameter - with the a phyllotactic pattern progressing from $[13,8,5]\rightarrow[5,3,2]\rightarrow[3,2,1]$ along the mast. Such a phenomenon can be described within the scheme developed here and a comparison ought to be made.

The framework described here is capable of variation in a number of ways. For example, it can be adjusted to allow the disks to grow in size. If disk growth were to keep pace with the increase of diameter, as a disk migrates down the stem, then the phyllotactic structure would not change - as in the case of ``decussate'' patterns \cite{Keith2007}.

Furthermore, a different choice of the initial (emerging) structure can produce variants. For example, if (instead of starting with the simplest possible structure - i.e. the $[1,1,0]$ arrangement) the emerging structure at the meristem is the $[2,2,0]$ arrangement then (applying the same rules as above) from Fig (\ref{fig3}) the progression of structures will be $[2,2,0]\rightarrow[4,2,2]\rightarrow[6,4,2]...$, which is indeed observed in some (so called \emph{whorled}) cases \cite{Keith2007}.  We shall explore these and other problems in future publications. 
 
\acknowledgments
The authors thank Alan Newell for stimulating discussions. D.W. acknowledges support from European Space Agency ESA MAP Metalfoam (AO-99-075) and Soft Matter Dynamics (AO-09-943+99-108+09-813). A. M. acknowledges support from the Aberystwyth University Research Fund.

\bibliographystyle{nonspacebib}

\begin{thebibliography}{10}

\bibitem{airy1872leaf}
H.~Airy, Proceedings of the Royal Society of London \textbf{21}, 176 (1872).

\bibitem{hofstadter2013alan}
D.~Hofstadter and C.~Teuscher, \emph{Alan Turing: Life and legacy of a great
  thinker} (Springer Science \& Business Media, 2013).

\bibitem{levitov1991fibonacci}
L.~Levitov, JETP letters \textbf{54}, 542 (1991).

\bibitem{levitov1991energetic}
L.~Levitov, EPL (Europhysics Letters) \textbf{14}, 533 (1991).

\bibitem{douady1992phyllotaxis}
S.~Douady and Y.~Couder, Physical Review Letters \textbf{68}, 2098 (1992).

\bibitem{atela2002dynamical}
P.~Atela, C.~Gol{\'e}, and S.~Hotton, Journal of Nonlinear Science \textbf{12},
  641 (2003).

\bibitem{Pennybacker:2013kp}
M.~Pennybacker and A.C. Newell, Physical Review Letters \textbf{110}, 248104
  (2013).

\bibitem{Mitchison:1977uy}
G.J. Mitchison, Science \textbf{196}, 270 (1977).

\bibitem{van1907mathematische}
G.~Van~Iterson, \emph{Mathematische und mikroskopisch-anatomische Studien
  {\"u}ber Blattstellungen: nebst Betrachtungen {\"u}ber den Schalenbau der
  Miliolinen}, Ph.D. thesis, TU Delft, Delft University of Technology (1907).

\bibitem{Mughal:2011tg}
A.~Mughal, H.~Chan, and D.~Weaire, Physical Review Letters \textbf{106}, 115704
  (2011).

\bibitem{Mughal:2012jd}
A.~Mughal, H.~Chan, D.~Weaire, and S.~Hutzler, Physical Review E \textbf{85},
  051305 (2012).

\bibitem{Mughal:2014bk}
A.~Mughal and D.~Weaire, Physical Review E \textbf{89}  042307 (2014).

\bibitem{movie}
  \url{https://youtu.be/gFKeOZTKpZM}.

\bibitem{Reinhardt:2003}
D.~Reinhardt, et al.  Nature \textbf{426}, 255 (2003)

\bibitem{Newell:2008}
A. C.~Newell, P.~ Shipman, and S.~Zhiying.  Journal of Theoretical Biology \textbf{251} 3 (2008): 421-439.

\bibitem{Pennybacker:2015}
M.~Pennybacker, P.~ Shipman, and A. C.~Newell.  Physica D: Nonlinear Phenomena \textbf{306} (2015): 48-81.

\bibitem{Rivier:2006}
N.~Rivier, J.-F.~Sadoc, and J.~Charvolin.  The European Physical Journal E \textbf{39} 1 (2016): 1-11.

\bibitem{Keith2007}
R.~Keith, ed  \emph{Handbook of plant science. Vol. 1} (John Wiley \& Sons, 2007).

\end{thebibliography}

\end{document}